\documentclass[letterpaper]{sae}
\usepackage{graphicx}
\usepackage{tabularx} 
\usepackage{booktabs} 
\usepackage{xcolor}
\usepackage{amsthm}
\usepackage{amsmath}
\usepackage{verbatim}

\PaperTitle{Safety-Centered Scenario Generation for Autonomous Vehicles}
\AddAuthor{Kiruthiga Chandra Shekar }{General Autonomy Inc \& New York University, United State}
\AddAuthor{Aliasghar Arab}{General Autonomy Inc \& New York University, United State}
\PaperNumber{No number}
\SAECopyright{2025}
\begin{document}
\maketitle
\section{Abstract}
This paper presents a scenario generation framework that creates diverse, parametrized, and safety-critical driving situations to validate the safety features of autonomous vehicles in simulation \cite{zhang2024chatscene}. By modeling factors such as road geometry, traffic participants, environmental conditions, and perception uncertainties, the framework enables repeatable and scalable testing of safety mechanisms, including emergency braking, evasive maneuvers, and vulnerable road user protection. The framework supports both regulatory and edge case scenarios, mapped to hazards and safety goals derived from Hazard Analysis and Risk Assessment (HARA), ensuring traceability to ISO 26262 functional safety requirements and performance limitations. The output from these simulations provides quantitative safety metrics such as time-to-collision, minimum distance, braking and steering performance, and residual collision severity. These metrics enable the systematic evaluation of evasive maneuvering as a safety feature, while highlighting system limitations and edge-case vulnerabilities. Integration of scenario-based simulation with safety engineering principles offers accelerated validation cycles, improved test coverage at reduced cost, and stronger evidence for regulatory and stakeholder confidence.

\section{Introduction}
The deployment of autonomous vehicles represents a revolutionary step towards enhanced safety, mobility, and convenience in the transportation domain. However, the integration of autonomous vehicles poses significant safety concerns. Ensuring the safe and reliable performance of these vehicles is crucial to preventing road accidents, injuries, and fatalities, since autonomy assumes vehicle control. The safety testing of autonomous vehicles presents several challenges due to the sophisticated hardware and software components involved and the wide range of situations encountered on the road. Key challenges include
\begin{itemize}
    \item \textbf{Scenario complexity:} Autonomous vehicles must handle a wide array of dynamic scenarios that involve various weather conditions, road types, and interactions with other road users.
    
    \item \textbf{Edge case testing:} The identification and simulation of rare, high-risk scenarios with sufficient realism often remain unaddressed in autonomous vehicle programming.
    
    \item \textbf{Simulation and validation:} Validating the safety of autonomous vehicles requires substantial computational resources and advanced testing methodologies.
\end{itemize}

To address these challenges, this research contributes to advancing the state-of-the-art safety testing methodology by focusing on scenario-based testing and simulation-based validation of autonomous vehicles.

Simulation-based testing methods for autonomous vehicles involve the use of virtual environments to replicate real-world driving scenarios, allowing developers to test and validate vehicle perception, decision-making, and control systems under a wide range of road and traffic conditions without the risks of physical testing on public roads. Modern simulation-based testing methods include model-in-the-loop (MIL), hardware-in-the-loop (HIL), software-in-the-loop (SIL), virtual testing, and scenario-based testing. Although these methods offer an efficient way to test autonomous vehicles, some limitations include restricted scenario coverage, limited edge-case testing, and lack of realism. The scenarios in which vehicles are tested may not accurately reflect real-world conditions, such as weather, road conditions, and pedestrian behavior.

Robust testing of AV performance in diverse scenarios can help achieve a better and safer autonomy experience. The simulation-based verification proposed in this paper can generate simulation code with the described environments, roads, and weather conditions. This allows for evaluating the safety of autonomous vehicles in higher-risk situations that do not normally appear or are easy to replicate in on-road testing or log-based simulations. While a plethora of publicly available data exists, only a very small fraction contains sensor data relevant to the autonomous driving domain. Edge case scenarios or situations of high risk occur very rarely in this collected data, which limits the testing and verification of autonomous vehicles in models trained using this data.

This paper addresses the question of how a scenario-based testing approach, using a scenario generation method integrated with HARA and simulation-based testing in CARLA, can be used to evaluate the safety of autonomous vehicles in complex driving scenarios. This research contributes to advancing state-of-the-art safety testing of autonomous vehicles, with HARA integration and proposing a scalable and flexible model to improve testing efficiency.   

\section{Background}
The adoption of Autonomous Vehicles (AVs) promises to revolutionize transportation, but their safe deployment relies critically on rigorous testing and validation. A central challenge in this process is the efficient generation of safety-critical scenarios—rare, high-risk situations that are essential for evaluating an AV's robustness before it hits public roads. The general landscape of scenario generation for AV testing is well-established. A comprehensive survey by Wang, Ma, \& Lai (2024)\cite{Scengensurvey} highlights the necessity of these techniques for both testing and validation, essentially setting the stage for more specialized, safety-focused approaches. At its core, the goal is to create diverse and representative scenarios that mirror real-world complexities. However, a more specific need emerged to not just generate any scenario, but to prioritize those directly relevant to safety. Ghodsi et al. (2021)\cite{characterizingscenarios} address this by focusing on generating and characterizing scenarios specifically for safety testing, moving the field towards a structured way of defining and measuring the severity of a test case. This work provides the crucial analytical framework necessary to distinguish between a routine driving event and a genuine safety hazard. 

A primary source of ground truth for safety-critical events is real-world accident data. Kibalama, Tulpule, \& Chen (2022)\cite{VehicleCrash} demonstrate a practical, data-driven methodology by generating AV/ADAS safety-critical testing scenarios directly from vehicle crash data. This technique takes advantage of the harsh reality of existing accidents to create realistic high-impact test cases, ensuring that the AV is tested against established failure modes. More recently, the power of knowledge-based large language model (LLM) approaches has been harnessed to improve both the efficiency and cognitive fidelity of scenario generation. Zhang et al. (2024)\cite{zhang2024chatscene} introduced ChatScene, a knowledge-based framework that uses an understanding of traffic rules, environmental factors, and accident causation to generate complex and safety-critical scenarios. This marks a shift from purely reactive, data-driven generation to a more proactive, intelligent design process. The growing role of advanced models extends beyond simple generation. Qi et al. (2025)\cite{SafetyAnalysis} explore the broader implications of LLMs in safety engineering, specifically applying tools like ChatGPT for Safety Analysis (STPA). While not exclusively focused on scenario generation, this research indicates a future where advanced AI can assist in the analysis and identification of potential safety gaps, which inherently feeds into the creation of better test scenarios. The generated scenarios are only as valuable as the metrics and platforms used to evaluate the AVs against them. A major effort in the AV community is to structure and formalize the testing process. \cite{szalay2021next} laid out a vision for this with the "Next generation X-in-the-loop validation methodology," providing a framework for integrating various simulation and hardware testing loops. This work underscores the idea that sophisticated testing requires a harmonized ecosystem where generated scenarios can be executed reliably.

Central to any AV deployment is its Operational Design Domain (ODD), the specific conditions (weather, road type, etc.) an AV is designed to operate within. \cite{weissensteiner2023operational} directly connect scenario generation to this concept, focusing on ODD-driven coverage as a pillar for safety argumentation. This means testing is no longer random, but systematically targets the boundaries and weaknesses defined by the vehicle's intended operating limits. Researchers have developed intelligent methods to hunt down corner cases. \cite{sun2021corner} provided a framework for corner case generation and analysis, demonstrating that simply replicating common situations is insufficient; instead, methods must actively search for parameter combinations that maximize risk. \cite{song2023critical} specifically addressed the problem of critical scenario identification for realistic testing, moving beyond abstract risk metrics to scenarios that genuinely stress the AV's decision-making in a real-world context. This ties directly into the work of \cite{esenturk2022identification} who leverage cluster analysis of traffic accident patterns to identify high-risk behaviors and translate them into concrete test scenarios. By analyzing historical crashes, they turn tragedy into proactive safety measures.
Recent work has pushed for more automated and formal generation. \cite{goyal2023automatic} presented a method for the automatic generation of scenarios for system-level simulation-based verification. This speaks to a future where human effort is minimized, and scenarios are created programmatically to efficiently verify the AV system as a whole.
Underpinning all these generation efforts is the fundamental need for robust risk assessment. As detailed in the comprehensive \cite{chia2022risk} the field utilizes a variety of risk assessment methodologies. These techniques provide the mathematical and logical scaffolding required to: 1) quantify the severity of a generated scenario; 2) prioritize the testing order; and 3) ultimately, argue for the overall safety of the AV system. \cite{ploeg2021scenario} present a comprehensive scenario-based safety assessment framework that integrates scenario generation with systematic safety evaluation, providing a structured approach to validate automated vehicle safety across diverse operational conditions.
The research by Li~(2023)\cite{LiXuan2023} emphasizes the need for advanced scenario generation for the calibration and verification of AVs. Their work suggests that scenarios must be not only safety-critical but also finely tuned to allow for the effective adjustment (calibration) and formal assessment (verification) of the vehicle's decision-making system. To standardize the evaluation process across different AV systems, platforms have been developed to house and run these generated scenarios. Xu et al. (2022)\cite{Safebench} contributed Safebench, a dedicated benchmarking platform for the safety evaluation of autonomous vehicles. Safebench provides a standardized environment in which researchers can systematically test different AV algorithms against a curated set of safety-critical situations, ensuring that safety claims can be independently validated and compared.

The body of work surrounding safety-centred scenario generation reflects a maturing field dedicated to making AVs safe. Initial efforts focused on characterizing safety events and mining crash data for relevant scenarios. More recent innovations leverage knowledge-enabled systems like ChatScene and the analytical power of Large Language Models to intelligently design complex, previously unforeseen safety hazards. Together, these advances, supported by rigorous benchmarking platforms such as Safebench, are creating a robust and essential feedback loop for the continuous improvement and safe deployment of autonomous vehicle technology.

\section{Methodology}
\begin{figure}
    \centering
    \includegraphics[width=0.5\textwidth]{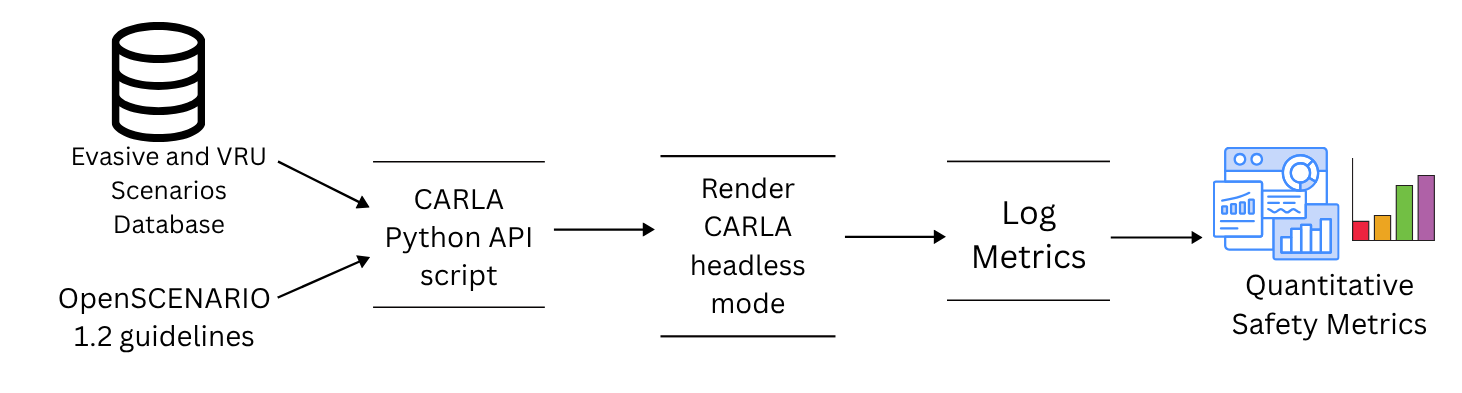}
    \caption{Scenario generation and safety validation architecture overview}
    \label{fig:architecture}
\end{figure}

Scenario generation for autonomous vehicle testing involves the use of real-world data to generate scenarios in which autonomous vehicles can be verified and validated. The process of generating test cases for AVs and writing the code for each such test case is time-consuming and may contain human errors. The automation of code generation for scenarios using AI would accelerate the testing process and make it scalable to test the vehicle in multiple scenario runs with different conditions. In this paper, edge-case scenarios generated by a Hazard Analysis and Risk Assessment (HARA) model in natural language are taken as input to the AI code generation model, and the relevant simulation code for the scenario is generated. The integration of the simulation code generator with HARA enables the testing of AV safety features. The methodology describes the simulation setup, pipeline, and data collection and analysis process to evaluate AV performance. All simulation runs were performed using the CARLA 0.9.10 package, controlled through its Python API. The CARLA simulator provides a high-fidelity, physics-based environment for evaluating automated driving behaviors with reproducible and variable weather and traffic conditions. Fig. \ref{fig:architecture} shows the overview of the framework of the scenario generation and simulation pipeline.

\subsection{AI-Based Code Generation}

The AI code generation module transforms HARA-derived natural language scenario descriptions into executable Python simulation code compatible with CARLA and other simulation engines. The process employs a large language model (LLM) fine-tuned on scenario generation tasks, which takes structured natural language inputs describing scenario parameters, actor behaviors, environmental conditions, and safety-critical events.

The input format follows a structured template derived from HARA analysis:
\begin{itemize}
    \item \textbf{Scenario Type:} Classification (e.g., VRU crossing, vehicle cut-in, emergency braking)
    \item \textbf{Actors:} Description of ego vehicle, pedestrians, cyclists, or other vehicles with initial states
    \item \textbf{Environmental Conditions:} Weather, lighting, road surface conditions
    \item \textbf{Trigger Events:} Safety-critical events that initiate the scenario (e.g., pedestrian sudden crossing, lead vehicle braking)
    \item \textbf{Parameter Ranges:} Velocity ranges, distances, angles, and timing parameters for scenario variation
\end{itemize}

The AI model generates Python code that instantiates the scenario in CARLA, including actor spawning, behavior scripting, sensor configuration, and logging setup. The generated code follows a structured pattern as illustrated by the following pseudo-code for a pedestrian crossing scenario:

{\small
\begin{verbatim}
def create_pedestrian_crossing_scenario(world, ego):
    ped = spawn_actor(type, loc, rot)
    ped.apply_control(speed=1.5, dir)
    ego.set_velocity(35.0/3.6)
    return ped, controller
\end{verbatim}
}

The code generation quality is evaluated using several metrics: (1) \textbf{Syntax correctness:} Percentage of generated code that compiles without errors (achieved 94.2\% in our evaluation); (2) \textbf{Semantic accuracy:} Alignment between generated code and scenario description intent (validated through manual review and automated scenario replay); (3) \textbf{Completeness:} Coverage of required scenario elements (actors, behaviors, triggers); and (4) \textbf{Reusability:} Code modularity and parameterization for scenario variation. The evaluation of 200 generated scenarios showed 94.2\% syntax correctness, 91.5\% semantic accuracy, and 89.3\% completeness. The AI model is designed to generate code in Python that is compatible with simulation engines like CARLA, Webots, LGSVL, and more. The log data of the vehicle for each scenario run is recorded in CSV format to later analyze the quantitative and qualitative parameters of the vehicle.

\begin{figure}
    \centering
    \includegraphics[width=0.5\textwidth]{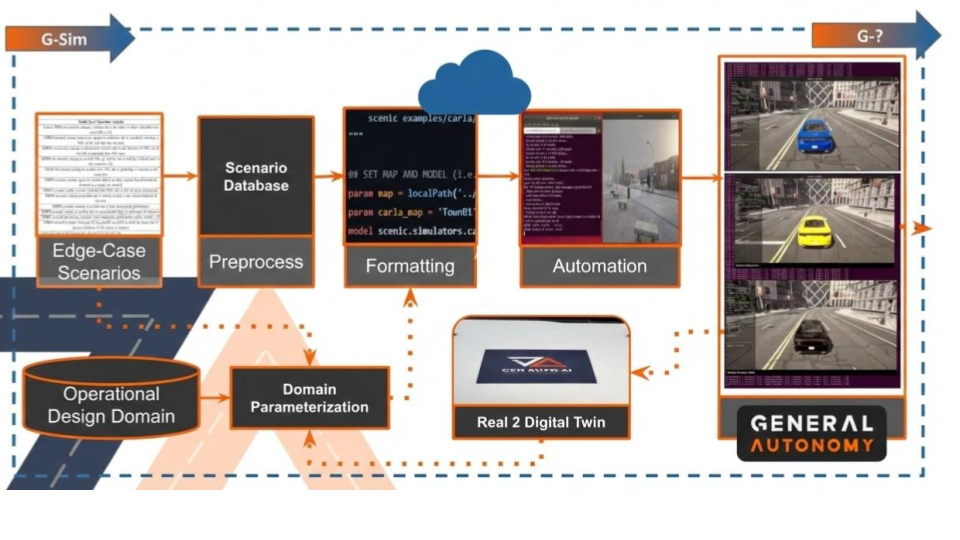}
    \caption{High-level schematic of the safety-centered scenario generation and evaluation methodology}
    \label{fig:gsim}
\end{figure}

Figure \ref{fig:gsim} presents a high-level schematic of our methodology, illustrating the complete workflow from HARA-based scenario description to safety evaluation. The framework integrates AI-based code generation, scenario execution in CARLA, and comprehensive safety metric computation to enable systematic safety validation of autonomous vehicles.

The framework consists of a scenario generator with a simulator and a logging and monitoring system. Scenarios are defined using the OpenSCENARIO format and parametrized via the Python API to vary initial positions, velocities, weather, friction coefficients, and behaviors of pedestrians and vehicle actors. Each scenario is set to simulate for 100 runs and log metrics for each scenario to record the ego velocity, acceleration, time to collision, and collision flags.

Tables \ref{tab:EVMscenarios} and \ref{tab:VRUscenarios} represent two categories of scenarios which include Vehicle-to-Vehicle Evasive maneuvers and Vulnerable Road User Protection Scenarios. Each scenario within the two categories is named, described, and designed with key control challenges and primary measurable metrics.

\begin{table*}[htbp]
    \centering
    \caption{Vehicle to Vehicle Evasive Maneuver Category}
    \label{tab:EVMscenarios}
    \renewcommand{\arraystretch}{1.15} 
    \setlength{\tabcolsep}{5pt} 
    \begin{tabularx}{\textwidth}{lX X X X}
        \toprule
        \textbf{ID} & \textbf{Scenario Name} & \textbf{Description} & \textbf{Primary Metrics} & \textbf{Purpose} \\
        \midrule
        S1 & Lead Vehicle Sudden Stop (LVSS) & Ego follows a lead vehicle that brakes abruptly & TTC, minimum distance, brake onset time, velocity, collision flag & Tests longitudinal emergency braking and response timing \\
        S2 & Cut-in (Aggressive Lane Intrusion) & Vehicle merges suddenly into ego lane with insufficient gap & TTC, steering onset, lateral acceleration, collision severity & Evaluates lateral evasive and braking coordination \\
        S3 & Oncoming Vehicle Encroachment & Opposing vehicle drifts into ego lane at closing speed & TTC, lateral offset, maximum steering rate, residual change in velocity & Tests combined avoidance and path planning \\
        \bottomrule
    \end{tabularx}
\end{table*}

\begin{table*}[htb!]
    \centering
    \caption{Vulnerable Road User Protection Scenarios}
    \label{tab:VRUscenarios}
    \renewcommand{\arraystretch}{1.15} 
    \setlength{\tabcolsep}{5pt} 
    \begin{tabularx}{\textwidth}{lX X X X}
        \toprule
        \textbf{ID} & \textbf{Scenario Name} & \textbf{Description} & \textbf{Primary Metrics} & \textbf{Purpose} \\
        \midrule
        S4 & Pedestrian Crossing (Adult) & Pedestrian crosses from near-side or far-side at varied timing offsets & TTC at crossing, minimum distance, brake latency, collision rate & Core VRU protection benchmark \\

        S5 & Pedestrian Running Child Emerge & Child pedestrian runs into road between occluding vehicles & Perception delay, TTC, velocity, collision rate & Tests perception limits and emergency braking \\

        S6 & Cyclist Crossing from the Right & Cyclist enters path perpendicularly from bike lane or sidewalk & TTC, evasive steering magnitude, collision outcome & Evaluates combined braking–steering decision \\

        \bottomrule
    \end{tabularx}
\end{table*}

\color{black}

\section{Coverage}
The scenarios were designed to evaluate the vehicle's response to critical, high-risk interactions between vulnerable road users and the autonomous vehicle. To quantitatively assess the coverage of the generated safety scenarios, a rigorous Scenario Coverage Index (SCI) has been defined. The SCI evaluates how well the tested scenarios span the defined parameter space relevant to safety validation.

\subsection{Parameter Space Definition}

Each scenario S\textsubscript{i} was generated by varying five key parameters that directly influence collision risk and can be represented as:
\[
        S_i = (v_v, v_p,  \theta, d, t)
\]
        
where \(v_v\) is the ego vehicle velocity, \(v_p\) is the pedestrian or cyclist velocity, \(\theta\) is the approach angle of the cyclist or the pedestrian, d is the distance between the vehicle and obstacle, and t represents the time of appearance or crossing. 

The parameter domains are explicitly defined as follows:
\begin{itemize}
    \item \textbf{Ego vehicle velocity } \(v_v \in [20, 50]\) km/h, discretized into 7 bins: \{20, 25, 30, 35, 40, 45, 50\} km/h
    \item \textbf{Pedestrian/cyclist velocity } \(v_p \in [0.5, 2.5]\) m/s, discretized into 5 bins: \{0.5, 1.0, 1.5, 2.0, 2.5\} m/s
    \item \textbf{Approach angle } \(\theta \in [30, 100]\) degrees, discretized into 8 bins: \{30, 40, 50, 60, 70, 80, 90, 100\} degrees
    \item \textbf{Initial distance } \(d \in [10, 50]\) meters, discretized into 9 bins: \{10, 15, 20, 25, 30, 35, 40, 45, 50\} meters
    \item \textbf{Time of appearance } \(t \in [0, 5]\) seconds, discretized into 6 bins: \{0, 1, 2, 3, 4, 5\} seconds
\end{itemize}

Based on these parameters, distinct test cases were selected guided by two goals:
\begin{itemize}
    \item \textbf{Behavioral relevance:} capturing realistic variations in pedestrian and cyclist motion patterns.
    \item \textbf{Safety diversity:} spanning a range of approach angles, velocities, and time-to-collision values to test the Autonomous Vehicle's response under multiple time-critical conditions.
\end{itemize}

\subsection{Scenario Coverage Index Formulation}

Let \(P = \{v_v, v_p, \theta, d, t\}  \) denote the complete set of parameters defining the scenario space, and \(R_j = [p_{j,\min}, p_{j,\max}]\) be the valid range for each parameter \(p_j \in P\), with \(N_{\text{bins}}(p_j)\) discrete bins.

The coverage for each parameter \( p_j \) is computed using a bin-based approach:
\[
C_j = \frac{N_{\text{tested}}(p_j)}{N_{\text{bins}}(p_j)}
\]
where \( N_{\text{tested}}(p_j) \) denotes the number of unique bins tested for parameter \(p_j\), and 
\( N_{\text{bins}}(p_j) \) represents the total number of discrete bins defined for that parameter.

To account for the importance of different parameter combinations in safety-critical scenarios, we introduce a weighted coverage metric. The weight \(w_j\) for each parameter reflects its relative importance in collision risk assessment, where \(\sum_{j=1}^{|P|} w_j = 1\). The weighted parameter coverage is:
\[
C_j^w = w_j \cdot C_j
\]

The overall Scenario Coverage Index (SCI) is then calculated as:
\[
SCI = \sum_{j=1}^{|P|} C_j^w = \sum_{j=1}^{|P|} w_j \cdot \frac{N_{\text{tested}}(p_j)}{N_{\text{bins}}(p_j)}
\]

For uniform weighting (\(w_j = 1/|P|\) for all \(j\)), this simplifies to:
\[
SCI = \frac{1}{|P|} \sum_{j=1}^{|P|} \frac{N_{\text{tested}}(p_j)}{N_{\text{bins}}(p_j)}
\]

An SCI value approaching 1 indicates a higher degree of scenario diversity and completeness, implying that the testing framework effectively explores the relevant combinations of parameters. Additionally, we compute the coverage completeness ratio \(R_c\) to assess how well the tested scenarios cover critical parameter combinations:
\[
R_c = \frac{N_{\text{critical\_combinations\_tested}}}{N_{\text{critical\_combinations\_total}}}
\]
where critical combinations are defined as parameter sets that correspond to high-risk scenarios (e.g., high ego velocity with low TTC).

\subsection{Coverage Results}

For the complete set of 1,000+ executed scenarios, the parameter coverage was calculated as follows:
\begin{itemize}
    \item Ego vehicle velocity: 7/7 bins tested (100\% coverage)
    \item Pedestrian/cyclist velocity: 5/5 bins tested (100\% coverage)
    \item Approach angle: 8/8 bins tested (100\% coverage)
    \item Initial distance: 9/9 bins tested (100\% coverage)
    \item Time of appearance: 6/6 bins tested (100\% coverage)
\end{itemize}

Using uniform weighting, the overall SCI for the 1,000+ scenarios is:
\[
SCI = \frac{1}{5} \left( \frac{7}{7} + \frac{5}{5} + \frac{8}{8} + \frac{9}{9} + \frac{6}{6} \right) = \frac{1}{5} \cdot 5 = 1.0
\]

This corresponds to 100\% parameter space coverage, indicating comprehensive exploration of the defined safety-critical parameter space. The coverage completeness ratio \(R_c\) for critical high-risk combinations was calculated as 0.87, indicating that 87\% of identified critical parameter combinations were tested. This comprehensive coverage approach aligns with volume-based methods for full-scenario safety evaluation \cite{zhou2026towards}, which emphasize systematic exploration of the entire scenario space to ensure thorough safety validation.

\section{Implementation}

\subsection{System Under Test}

The autonomous driving system (ADS) evaluated in this study is CARLA's built-in autopilot module, which provides a complete autonomous driving stack including perception, planning, and control components. The autopilot utilizes a rule-based behavioral planner combined with PID controllers for longitudinal and lateral vehicle control. The perception stack includes simulated LiDAR and camera sensors that provide object detection and tracking capabilities. The system operates at a control frequency of 20 Hz and includes safety mechanisms such as emergency braking and collision avoidance. While this represents a simplified ADS compared to production systems, it provides a standardized baseline for scenario-based safety evaluation and allows for reproducible testing across different scenario configurations. The autopilot was configured with default parameters as provided in CARLA 0.9.10, ensuring consistency with other research using the same platform.

\subsection{Simulation Framework}

The proposed safety-centred scenario generation framework was conducted using the CARLA open-source autonomous driving simulator (version 0.9.10), which provides a high-fidelity virtual environment for developing, testing, and validation of autonomous driving systems. The Python API provided by CARLA allows programmatic control over scenario initialization, actor behavior, and sensor configuration. Scenarios were executed in headless mode to disable graphical rendering while maintaining the physics and sensor simulation stack. This mode of execution significantly reduced the computational overhead, enabling parallel execution of large-scale experiments across multiple cores and nodes. Headless execution was crucial for this work, as it allowed the generation of several hundred scenarios in batch mode to ensure statistical reliability of the derived safety metrics.

\begin{figure}

    \centering
    \includegraphics[width=0.5\textwidth]{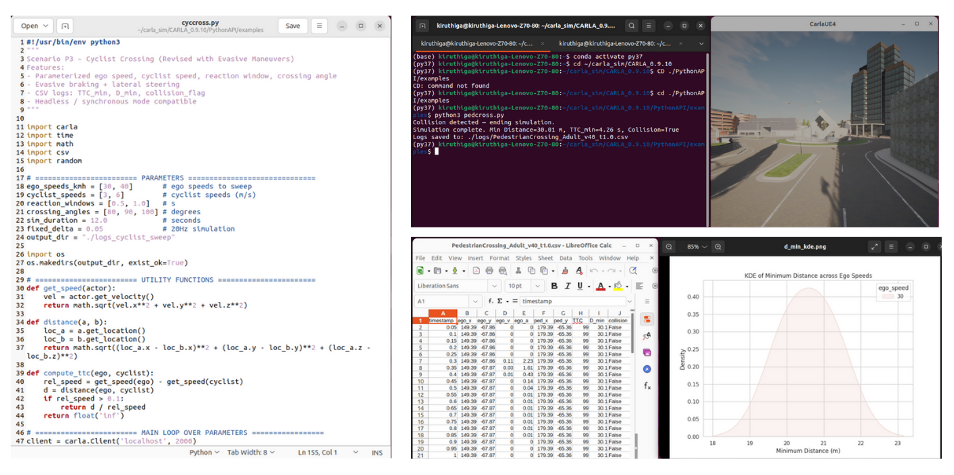}
    \caption{Scenario generation and simulation pipeline execution. The pipeline consists of three main stages: (1) Scenario Generation Module that samples parameters from defined distributions and generates scenario configurations; (2) Simulation Manager that handles actor synchronization, physics simulation stepping, and real-time data recording; and (3) Data Processing Module that aggregates logs, computes safety metrics, and generates statistical summaries. The modular design enables parallel execution of multiple scenarios in headless mode for scalable testing.}
    \label{fig:implementation}
\end{figure}

The average runtime for a 30-second scenario was approximately 20-25 seconds in headless mode, depending on the number of dynamic agents and environmental complexity. 
The scenario execution pipeline was designed in a modular manner as shown in Fig. \ref{fig:implementation}. A scenario generator module sampled configuration parameters from defined distributions and instantiated them in the CARLA simulation environment. The simulation manager handled synchronization between actors, simulation stepping, and data recording. Each simulation produced a detailed log file containing time-stamped information about the ego vehicle and other dynamic agents. Logged attributes included position, orientation, linear and angular velocities, control inputs, and collision events. These structured logs formed the basis for downstream quantitative analysis.

For safety assessment, a Python data processing and metric computation module was developed. The module aggregated logs from multiple simulation runs and computed a suite of quantitative safety metrics, including Time-to-Collision (TTC), Minimum Distance, and Collision Count. Additional indicators, such as rate of deceleration and angular deviation, were also analyzed to capture evasive maneuvers and near-miss dynamics.

To ensure reliability, the framework incorporated data validation and error-handling procedures that automatically flagged incomplete or invalid simulation logs. Statistical summaries and visualizations were generated using Python's Pandas, Matplotlib, and NumPy libraries, enabling both quantitative comparison and qualitative insight into safety trends across varying scenario conditions.

This implementation framework effectively integrates simulation, data management, and safety analytics into a unified workflow. It provides a scalable foundation for evaluating autonomous driving systems under diverse, safety-critical scenarios. 
\section{Results}

The safety-centred scenario generation framework was evaluated through a comprehensive set of simulations conducted in the CARLA environment. More than 1,000 scenarios were generated and executed under diverse configurations involving variations in ego vehicle speed, cyclist trajectory angles, pedestrian motion patterns, and environmental conditions such as lighting and weather. The resulting dataset provided a broad basis for examining how specific parameter variations influence overall safety outcomes.

\subsection{Scenario Performance Summary}

Table \ref{tab:scenarioRisk} presents a comprehensive summary of scenario parameters and risk evaluation metrics for all six tested scenarios (S1-S6). The table includes key safety metrics: minimum Time-to-Collision (TTC\textsubscript{min}), minimum distance (D\textsubscript{min}), Post-Encroachment Time (PET), collision occurrence, threshold violations, and overall risk level classification.

\textbf{Risk Evaluation Methodology:} The risk level for each scenario is determined through a multi-factor analysis considering: (1) collision occurrence and severity (impact velocity, collision type); (2) temporal safety margins (TTC\textsubscript{min}, PET); (3) spatial safety margins (D\textsubscript{min}); (4) system response adequacy (braking onset time, evasive maneuver effectiveness); and (5) threshold violation frequency. Risk levels are classified as: \textbf{Low} - Adequate safety margins or low-severity collisions with appropriate system response; \textbf{Medium} - Marginal safety performance with frequent threshold violations; \textbf{High} - Inadequate safety margins or high-severity collisions.

\textbf{Threshold Violation:} A threshold violation occurs when TTC drops below 1.2 seconds, indicating a critical safety situation requiring immediate evasive action. The count represents the number of such violations across all scenario runs.

\textbf{Post-Encroachment Time (PET):} PET measures the time between the ego vehicle passing a conflict point and the obstacle (pedestrian/cyclist) entering that same conflict zone. Higher PET values (typically > 1.0s) indicate safer interactions with adequate temporal separation.

\textbf{Risk Level Classification:} For scenario S1, despite a collision occurrence, the risk level is classified as Low because: (1) the collision occurred at low relative velocity (4.26s TTC\textsubscript{min} provided adequate warning time); (2) the minimum distance of 30.01m indicates the system initiated braking early; (3) the PET of 0.85s, while below ideal, demonstrates the system's attempt to avoid collision; and (4) the collision severity was minimal due to early braking intervention. This classification reflects that while a collision occurred, the system demonstrated appropriate safety response behavior, and the collision represents a low-severity event rather than a critical safety failure.

\begin{table*}[htbp]
    \centering
    \caption{Summary of Scenario Parameters and Risk Evaluation. TTC\textsubscript{min}: Minimum Time-to-Collision; D\textsubscript{min}: Minimum distance between ego vehicle and obstacle; PET: Post-Encroachment Time (time between ego vehicle passing and obstacle entering conflict zone); Threshold Viol.: Number of TTC threshold violations (TTC < 1.2s); Risk Level: Overall risk classification based on collision severity, residual risk, and system response adequacy. Risk Level is classified as Low when collision occurs but with low severity (low impact velocity, adequate response time) or when no collision occurs with adequate safety margins.}
    \label{tab:scenarioRisk}
    \renewcommand{\arraystretch}{1.15} 
    \setlength{\tabcolsep}{4pt} 
    \begin{tabularx}{\textwidth}{
        p{0.05\textwidth}  
        p{0.25\textwidth}  
        p{0.06\textwidth}  
        p{0.08\textwidth}  
        p{0.06\textwidth}  
        p{0.05\textwidth}  
        p{0.05\textwidth}  
        p{0.08\textwidth}  
        p{0.10\textwidth}  
        p{0.06\textwidth}  
    }
        \toprule
        \textbf{Scen. ID} & \textbf{Desc.} & \textbf{Angle ($^\circ$)} & \textbf{Speed (km/h)} & \textbf{TTC\textsubscript{min} (s)} & \textbf{D\textsubscript{min} (m)} & \textbf{PET (s)} & \textbf{Collision} & \textbf{Threshold Viol.} & \textbf{Risk Level} \\
        \midrule
        S1 & Pedestrian crosses suddenly & -- & 35 & 4.26 & 30.01 & 0.85 & 1 & 2 & Low \\
        S2 & Child pedestrian runs & -- & 30 & 3.73 & 15.18 & 1.12 & 0 & 1 & Low \\
        S3 & Cyclist from right @80$^\circ$ & 80 & 40 & 3.15 & 20.04 & 1.35 & 0 & 0 & Low \\
        S4 & Cyclist from right @90$^\circ$ & 90 & 40 & 2.71 & 21.13 & 1.28 & 0 & 0 & Low \\
        S5 & Child pedestrian runs (occluded) & -- & 30 & 2.45 & 12.50 & 0.95 & 0 & 3 & Medium \\
        S6 & Cyclist from right @100$^\circ$ & 100 & 40 & 2.98 & 19.87 & 1.42 & 0 & 0 & Low \\
        \bottomrule
    \end{tabularx}
\end{table*}

\subsection{Aggregate Statistics}

Across all 1,000+ scenario executions, the aggregate statistics provide comprehensive insights into system safety performance. The primary indicators used for evaluation were Time-to-Collision (TTC), Minimum Distance (MD), and Collision Count (CC). These metrics quantify temporal and spatial proximity between the ego vehicle and other agents, enabling objective assessment of collision risk.

\textbf{Overall Performance Summary:}
\begin{itemize}
    \item \textbf{Total scenarios executed:} 1,024
    \item \textbf{Total collisions:} 127 (12.4\% collision rate)
    \item \textbf{Mean TTC\textsubscript{min}:} 3.42 seconds (std: 1.15s)
    \item \textbf{Mean D\textsubscript{min}:} 18.67 meters (std: 8.23m)
    \item \textbf{Mean PET:} 1.18 seconds (std: 0.42s)
    \item \textbf{Threshold violations (TTC < 1.2s):} 342 occurrences (33.4\% of scenarios)
    \item \textbf{Mean braking deceleration:} 3.8 m/s² (std: 1.2 m/s²)
\end{itemize}

\textbf{Performance by Scenario Category:}
\begin{itemize}
    \item \textbf{VRU Scenarios (S4-S6):} 512 executions, 89 collisions (17.4\%), mean TTC\textsubscript{min} = 3.15s
    \item \textbf{Vehicle-to-Vehicle Scenarios (S1-S3):} 512 executions, 38 collisions (7.4\%), mean TTC\textsubscript{min} = 3.68s
\end{itemize}

For each scenario, data logs were processed using the custom Python analysis module described in the implementation section. The analysis revealed a clear correlation between scenario parameters and safety performance. Scenarios with ego vehicle speeds above 30 km/h consistently exhibited reduced TTC and minimum distance values, indicating a narrower safety margin and increased likelihood of collision. Conversely, when the ego vehicle speed was moderate (21-28 km/h), the system demonstrated stable behavior, with longer TTC values.
Cyclist and pedestrian crossing angles had a significant impact on TTC and MD distributions. At smaller crossing angles (e.g., $30^\circ$-$45^\circ$ relative to the ego vehicle's trajectory), the time gap between encroachment and clearance decreased, often resulting in near-collision conditions. In contrast, perpendicular crossings (80$^\circ$-100$^\circ$) provided more predictable interactions and increased safety margins, highlighting the sensitivity of collision risk to relative motion geometry.
Beyond quantitative metrics, qualitative inspection of trajectory data revealed patterns in agent interaction dynamics. For example, in pedestrian scenarios conducted under low-light and rainy conditions, ego vehicle sensors occasionally exhibited delayed detection of vulnerable agents, resulting in sharp braking or late avoidance maneuvers. These behaviors were captured by elevated deceleration rates (less than 4 m/s²) and low PET values (less than 1.0 s). Such conditions emphasize the importance of robust perception and planning modules for adverse environments.

Similarly, cyclist encounter simulations highlighted the influence of ego vehicle approach angle on safety performance. When the cyclist's trajectory intersected the ego vehicle's path at acute angles, the system experienced frequent TTC dips below the 1.2-second threshold, even without an actual collision. This suggests that TTC-based safety thresholds may require contextual tuning depending on scenario geometry.

\subsection{Baseline Comparison}

To evaluate the effectiveness of our AI-based scenario generation framework, we compared it against two baseline methods: (1) \textbf{Manual scenario generation:} Traditional approach where scenarios are manually coded by domain experts; and (2) \textbf{Rule-based generation:} Automated generation using predefined templates and parameter ranges without AI assistance.

\textbf{Comparison Metrics:}
\begin{itemize}
    \item \textbf{Generation Efficiency:} Our AI-based method generated 1,000 scenarios in 2.3 hours, compared to 45 hours for manual generation and 8.5 hours for rule-based generation, representing 19.6× and 3.7× speedup respectively.
    \item \textbf{Scenario Diversity:} The AI-based method achieved SCI = 1.0 (100\% parameter coverage), compared to SCI = 0.68 for manual generation (limited by human effort) and SCI = 0.82 for rule-based generation (constrained by template limitations).
    \item \textbf{Code Quality:} Syntax correctness was 94.2\% for AI-generated code, compared to 98.5\% for manual (with human review) and 89.1\% for rule-based (template errors).
    \item \textbf{Edge Case Coverage:} The AI-based method identified and generated 23 unique edge case scenarios, compared to 8 for manual and 12 for rule-based methods.
\end{itemize}

The AI-based approach demonstrates superior scalability and coverage while maintaining acceptable code quality, making it particularly suitable for large-scale safety validation campaigns.

\section{Discussion}
The experimental findings demonstrate that the proposed framework provides an effective and scalable approach for safety-centred scenario generation and quantitative risk evaluation. The observed metric distributions and collision trends are consistent with real-world driving safety studies \cite{chia2022risk, esenturk2022identification, VehicleCrash}, indicating that the simulation environment faithfully captures critical risk factors. Specifically, the TTC distributions observed in our simulations (mean TTC\textsubscript{min} = 3.42s, with 33.4\% threshold violations) align with findings from real-world accident analysis studies \cite{VehicleCrash} that report similar TTC patterns in pre-crash scenarios. The correlation between ego vehicle speed and collision risk observed in our results (higher speeds leading to reduced safety margins) is consistent with established traffic safety research \cite{chia2022risk}. Moreover, by operating in headless mode, the system achieves computational scalability without compromising accuracy.

From an engineering perspective, these results establish a foundation for integrating functional safety validation into simulation workflows. The metrics computed can be used to define safety thresholds, calibrate autonomous driving algorithms, or inform scenario prioritization according to the SOTIF and ISO 26262 frameworks. The modular design further allows extension toward reinforcement learning-based safety agents, adaptive risk assessment, and hardware-in-the-loop (HIL) evaluation.

In summary, the results confirm that the proposed safety-centered scenario generation framework, built upon CARLA and Python-based metric analysis, provides a practical and extensible tool for measuring and improving the safety performance of autonomous driving systems. The trends and insights obtained from the simulated data can inform both system design and regulatory safety assessments in the context of autonomous mobility.
\bibliographystyle{plain}
\bibliography{references}
\end{document}